\documentclass[apj]{emulateapj}
   
\def\Msun{M_\odot}

\newcommand{\appropto}{\mathrel{\vcenter{
  \offinterlineskip\halign{\hfil$##$\cr
    \propto\cr\noalign{\kern2pt}\sim\cr\noalign{\kern-2pt}}}}}
\usepackage{pslatex}
\usepackage{graphicx}
\usepackage[flushleft]{threeparttable}

\shorttitle{X-ray surface brightness profiles of optical AGN}
\shortauthors{Mukherjee et al.}

\begin{document}
\title{X-ray Surface Brightness Profiles of Optically Selected Active Galactic Nuclei: Comparison with X-ray AGN} 
\author{Sagnick Mukherjee$^{1}$, Anirban Bhattacharjee$^{2,3}$, Suchetana Chatterjee$^{1}$, Jeffrey A.\ Newman$^{4,5}$, Renbin Yan$^{6}$ }
\affiliation{$^{1}${Department of Physics, Presidency University, Kolkata, 700073, India},\\ $^{2}${Department of Physics and Astronomy, University of Wyoming, Laramie, WY 82072, USA},\\$^{3}${Department of Biology, Geology and Physical Sciences, Sul Ross State University, East Highway 90
Alpine, TX 79832, USA}, \\$^{4}${Department of Physics and Astronomy, University of Pittsburgh, Pittsburgh, PA 15260, USA}, \\$^{5}${PITT-PACC, University of Pittsburgh, Pittsburgh, PA 15260, USA},\\$^{6}${Department of Physics and Astronomy, University of Kentucky, Lexington, KY 40506, USA} }

\email{sagnickm@yahoo.in}
\email{suchetana.physics@presiuniv.ac.in}

\begin{abstract}
We use data from the All Wavelength Extended Groth Strip International Survey to construct stacked X-ray maps of optically bright active galaxies (AGN) and an associated control sample of galaxies at high redshift ($z \sim 0.6$). From our analysis of the surface brightness profiles obtained from these X-ray maps, we find evidence of feedback from the active nuclei. We find that excluding galaxies and AGN, residing in group environments, from our samples enhances the significance of our detection. Our results support the tentative findings of Chatterjee et al. who use X-ray selected AGN for their analysis. We discuss the implications of these results in the context of quantifying AGN feedback and show that the current method can be used to extract X-ray source population in high redshift galaxies.

\end{abstract}

\section{Introduction}

The influence of the central supermassive black hole (SMBH) on the growth and evolution of the host galaxy is evident through a variety of observational and theoretical investigations \citep[e.g.,][]{s&r98,gebhardtetal00, m&f01,Dimatteoetal2005Natur, p&f06,mcnamaranulsen07, m&n07,chatterjeeetal08, chatterjeeetal10,  battagliaetal10,      gittietal12, pelligrinietal12, gasparietal12, choietal13,chatterjee15}. The effect of the central SMBH on large scale gas and stars in galaxies and clusters is generally known as active galactic nuclei (AGN) feedback in the literature. 

The first evidences of interaction between the gas and stars in galaxies and clusters with the central AGN came through pioneering X-ray and radio observations of galaxy clusters \citep[e.g.,][]{birzan04,nulsen05,dunnfabian06}. These pathbreaking observations gave notion to the theory of jet-intra cluster medium (ICM) interaction \citep[e.g., ][]{cr98,cr01,cr102,cr02,choi04,cr105,cr05,fab05,vernaleo06,vernaleo07,cr09,cr10,cr12,cr15,yang16,clark16,
yang2016,yang17}. The jet-ICM interaction was also observed in group environments \citep{komossa99,vrtilek00,zanni05,giodini10,david11,Randall11,src11,randall14,
fertig14,verdes15,randall15}. Recently efforts were made by \citealt{chatterjee15} (C15 hereafter) to observe the AGN-ICM/ISM interaction in galaxy environments. 

Using data from the All Wavelength Extended Groth Strip International Survey (AEGIS) \citep{davisetal07} and the DEEP-2 galaxy redshift survey \citep{davis03,newman13} C15 used AGN and galaxy samples ($0.3 \leq z \leq 1.3$), matched on the basis of their optical properties (e.g., colour and redshift) and extracted their X-ray surface brightness profiles. The results were suggestive of an excess in the extended X-ray emission of normal galaxies to that of AGN host galaxies. However the effect of the {\it Chandra} point spread function (PSF) of the X-ray bright AGN sample left the conclusion tentative. In this work we follow the same technique as C15 but carry out the analysis with an optically selected sample \citep{renbin11} to minimize the effect of the X-ray emission from the AGN itself. Our results support the findings of C15 and we propose that even at galaxy environments we find observational evidences of interaction between the central AGN and large scale gas in the galaxy.

Our paper is organized as follows. In \S 2 we give a brief description of our datasets and methodology. We present our results in \S 3. Finally we summarize and discuss our results in \S 4. Throughout the paper we assume a spatially flat, $\Lambda$CDM cosmology: $\Omega_{m}=0.28$, $\Omega_{\Lambda}=0.72$, $\Omega_{b}=0.04$, and $h=0.71$ \citep{spergel07}.

 \section{Data sets and Methodology}
We use data products from the AEGIS and the DEEP-2 galaxy redshift survey for our analysis. 

\subsection{Optically selected AGN and Galaxy Control Samples}

The AGN source data set contains $227$ optically selected AGN sources. The optical selection of the AGN is carried out using the $\left[O_{III} \right ]/H_{\beta}$ line ratio and the $U-B$ rest frame color \citep{renbin11}. Out of the total $291$ AGN, selected via the optical line ratios, $64$ of them are X-ray detected \citep{laird09} and the rest $227$ are purely optical AGN candidates with no X-ray counterparts. Typically AGN selection in the nearby Universe (z \textless 0.4) is done using optical emission line ratios of $\left[O_{III} \right ]/H_{\beta}$ and $\left[N_{II}\right]/H_{\alpha}$ or by their X-ray fluxes. These line ratios are used to classify AGN and star forming galaxies in the local universe \citep[e.g.,][]{kewley01,kauffman03} using the BPT diagram \citep{bpt81}. But for higher redshift sources, the $N_{II}$ and $H_{\alpha}$ lines redshift out of the optical region into near infrared. So the use of new BPT diagrams with $\left[O_{III} \right ]/H_{\beta}$ and $U-B$ colour has been proposed in the literature \citep[e.g.,][]{renbin11}.

\begin{figure*}[t]
\begin{center}
\begin{tabular}{c}
        \resizebox{8cm}{!}{\includegraphics{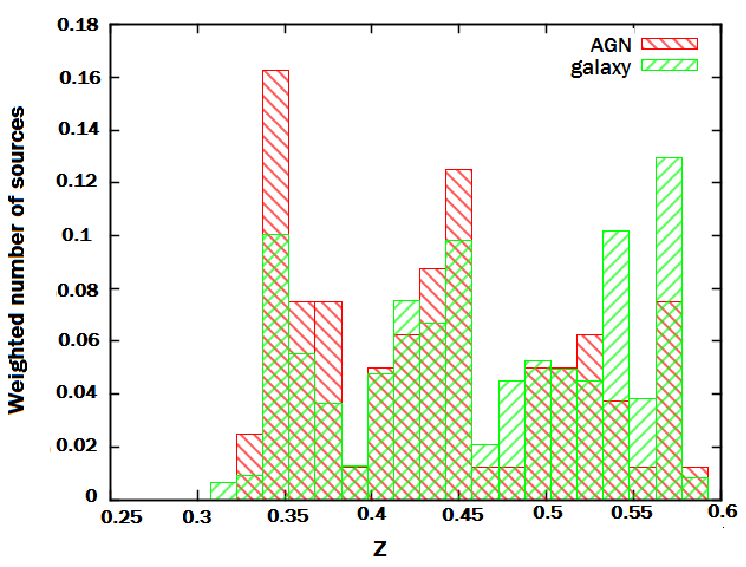}}
        \resizebox{8cm}{!}{\includegraphics{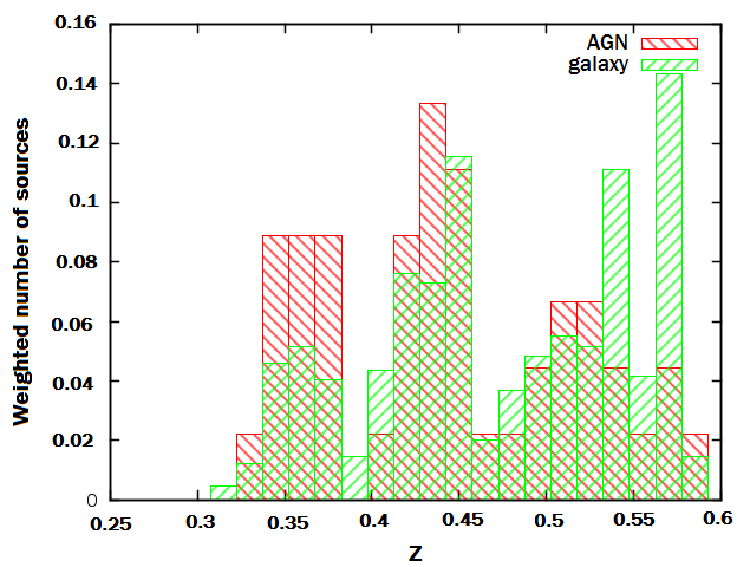}}
              
          \end{tabular}
 \caption{Redshift distribution of the complete AGN and galaxy samples (left panel) and the `isolated' AGN and galaxy samples (right panel). See \S 2 and Table 1 for description of the samples. }
\end{center}
\label{zdist}
\end{figure*}

The other method of classification that is used for selecting AGN is the luminosity cut in X-rays (L$_{2-10 keV}> 10^{42}$ erg s$^{-1}$) as no star-forming galaxies in the recent Universe is seen to have X-ray brightness higher than this limit \citep[e.g.,][]{ranalli03,renbin11}. Our optically selected sources are re-classified as AGN by using the X-ray cut. In our sample 76\% of the optically selected AGN candidates are not classified as X-ray AGN (by using the luminosity cut). To confirm that the optically selected and X-ray rejected AGN candidates are not dusty high redshift star-forming galaxies, the stellar mass distribution of the galaxies are considered. The stellar mass contents of our AGN candidates are found to be identical to the X-ray selected AGN and also they are much more massive than star-forming galaxies \citep{renbin11}. We note that there might be some AGN in our sample with lower X-ray luminosities that can appear due to lower accretion rates or intrinsic high X-ray absorption of these sources. These sources would not be classified as AGN according to the criterion of \citet{renbin11}. However, the existence of those undetected X-ray sources will not have any effect on our analysis, because we want to exclude the sources that have been classified as bright X-ray point sources and hence minimize the PSF contamination. 

\begin{table}[t]
\begin{center}
\begin{tabular}{c|c|c} 
 \hline
\multicolumn{1}{c|}{Selection applied to Sample }&
\multicolumn{1}{c|}{Number of AGN}&
\multicolumn{1}{c}{Number of Galaxies} \\
 \hline\hline
  Initial sample & 227 & 5000 \\ 
 \hline
  $0.3 \leq z \leq 0.6$  & 80 & 1576 \\
  \hline
  Group Catalog & 45 & 892 \\
  \hline\hline
 \end{tabular}
\caption{Table showing the various cuts applied to the initial sample of 5000 galaxies and 227 optically selected AGN. The sample is reduced to 1576 galaxies and 80 AGN after the redshift cut. This is our `complete sample'. After removing the group belonging sources we are left with 45 AGN and 892 galaxies which are our `isolated sample'. }
\end{center}
\end{table}

To compare the X-ray environments of our optically selected AGN candidates with that of galaxies we need to construct an equivalent sample of normal galaxies. To do this, we construct a control sample of galaxies by defining a `matching criterion' using a three-dimensional parameter space involving B-band absolute magnitude (proxy for luminosity), $U-B$ color and redshift with our AGN sample. The normal galaxy control sample contains a total of $5000$ galaxies. We want the host galaxies of the AGN and the control sample to be identical in their properties in order to ensure that whatever impact we see in our analysis comes from the black-hole activity and not from the stellar population or other constituents of the host galaxies. See \S 4 for more discussion on this issue. $U-B$ color matching has been done in order to ensure that the AGN host and normal galaxies have similar stellar population and star-formation rates (C15). For our analysis, the X-ray classified AGN (L$_{2-10 keV}> 10^{42}$ erg s$^{-1}$) and optically classified AGN are filtered out from the galaxy control sample.
\begin{figure*}[t]
\begin{center}
\begin{tabular}{c}
        \resizebox{8cm}{!}{\includegraphics{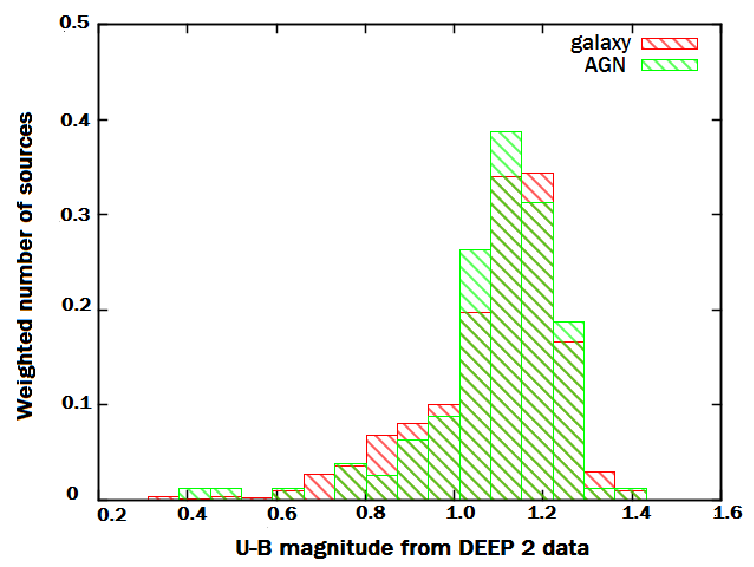}}
        \resizebox{8cm}{!}{\includegraphics{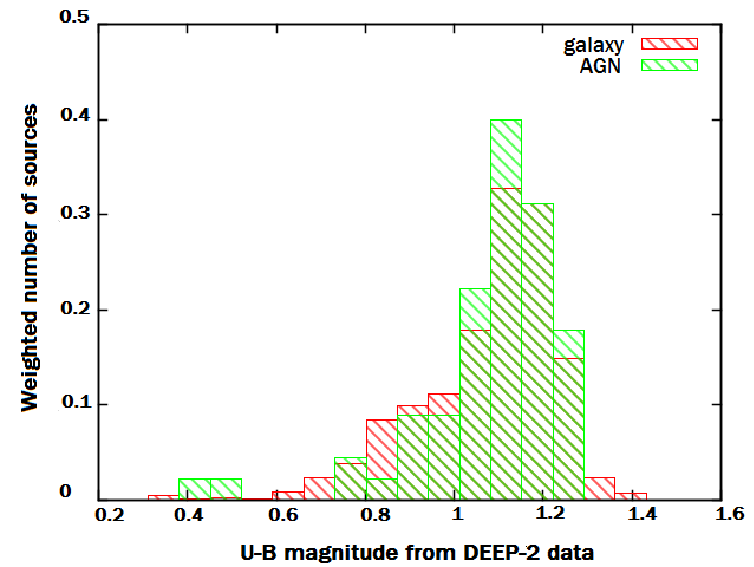}}      
                  
          \end{tabular}
 \caption{$U-B$ colour distributions of the the complete AGN and galaxy samples (left panel)  and the `isolated' AGN samples (right panel). See \S 2 and Table 1 for description of the samples.}
\end{center}
\label{Fig.2}
\end{figure*}

It is also seen in the literature that the galaxy properties depend on redshifts \citep[e.g.,][]{beifori14,lapi14} and hence the samples (AGN and control) are matched in redshift too. Amongst the $5000$ galaxies, 1673 galaxies have single entries only, while 1088 entries have been repeated multiple times  in the control sample in order to make the color and redshift distribution of the two samples as similar as possible. The distributions are shown in the left panel of Fig.\ 1. A K-S test ensured a significance p value of $0.97$.

\begin{figure*}[t]
\begin{center}
\begin{tabular}{c}
 \resizebox{4cm}{!}{\includegraphics{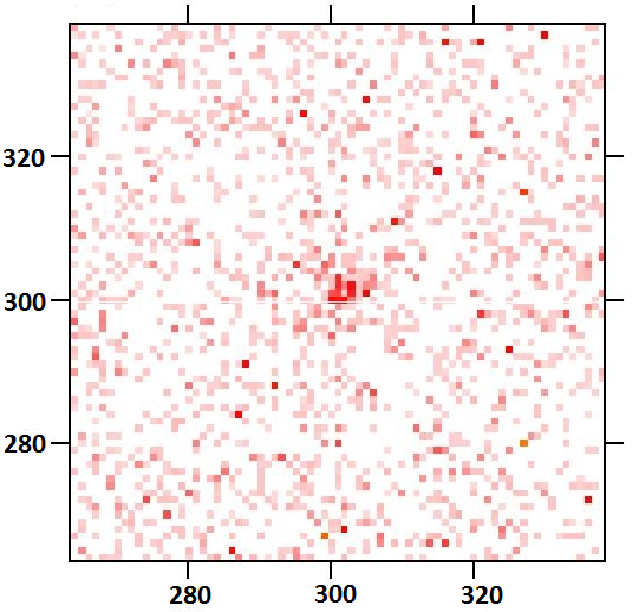}}
 \resizebox{4cm}{!}{\includegraphics{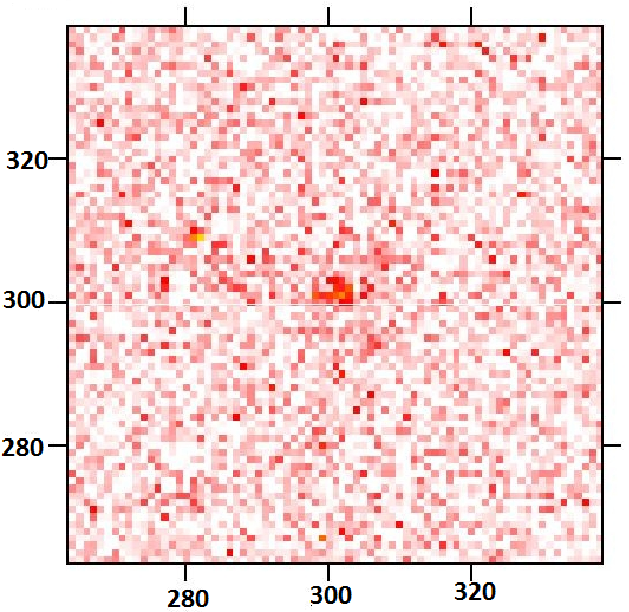}}
 \resizebox{4cm}{!}{\includegraphics{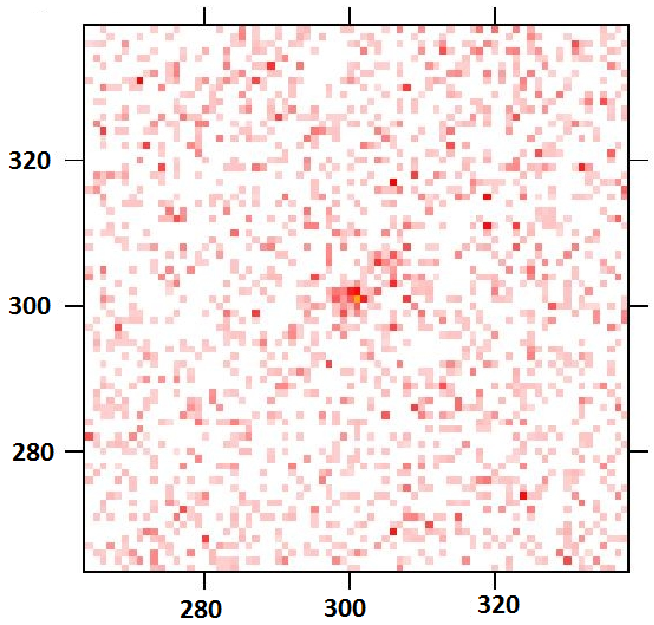}}
 \resizebox{4cm}{!}{\includegraphics{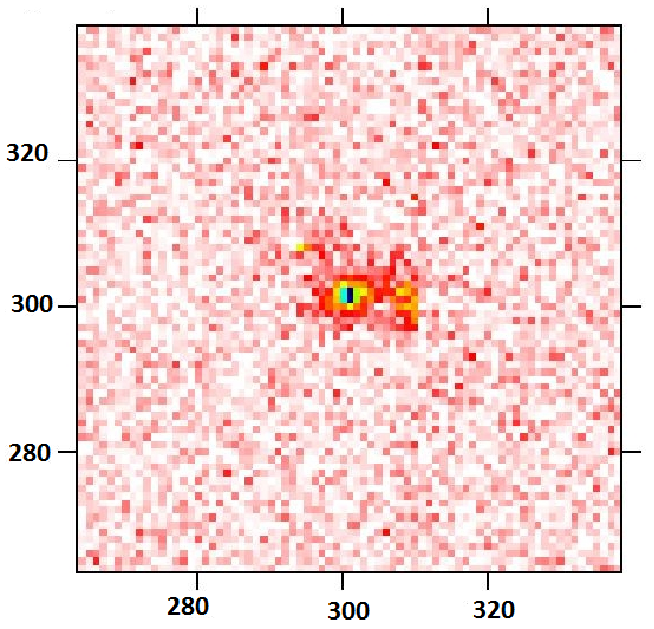}}
 \end{tabular}
\caption{Stacked maps corresponding to the four samples shown in Fig.\ 1 Fig.\ 2 and Table 1. `Complete AGN sample' (left-most panel), `Complete galaxy sample' (middle left panel), `isolated AGN sample' (middle right panel) \& `isolated galaxy sample' (right-most  panel). The original dimension of the stacked maps are of $300''x300''$ but only a $40''x40''$ square region around the centre has been shown here. See \S 2.3 for the stacking methodology.}
\end{center}
\label{Fig.3}
\end{figure*}

\subsection{Group Catalog}
It is known that group environments are likely to affect the extended X-ray emission due to the enhanced depth of the gravitational potential well. Hence to identify the group environments associated with the AGN and galaxies in our sample we use the DEEP2 group catalog \citep{gerke12}. \citet{gerke12} used the Voronoi-Delaunay method for finding groups of galaxies in the redshift space. Two or more galaxies are said to belong to a group if they are linked by the group finding algorithm. Galaxies which are not linked by the group finder algorithm are termed as isolated or field galaxies. The catalog enlists the group ID of $34,492$ DEEP2 sources and those having group ID's greater than $10000$ are isolated (see \citealt{gerke12} for details). After eliminating the sources belonging to galaxy groups we obtain $45$ AGN and $892$ galaxy sources from our parent sample. We identified these sources as `isolated'. We also examine the redshift distribution of these two `isolated' samples in order to compare them with the complete sample. The redshift distribution of the samples are shown in the right panel of Fig.\ 1. We perform a K-S test on the redshift distribution of the `isolated' galaxy and AGN samples and obtain a p-value of $0.97279$ indicating sufficient preservation of similarity. The color distribution of the samples are shown in Fig.\ 2. The KS-test when performed on the color distribution yielded a p value of $0.88275$. In Table 1 we provide the details of the constructed samples used in our analysis.

\subsection{Methodology}
We follow the stacking analysis technique adopted in C15. The effective area-exposure corrected (EAEC maps) X-ray maps of AEGIS-X \citet{laird09} have been used for the preparation of the stacked images of our samples. These maps are constructed using the event files, exposure time maps and effective area maps provided in \citep{laird09}. A region with dimensions of $5\times5$ arcminutes has been mapped around each of our identified sources from the EAEC maps. All the other sources present in this region were identified using the AEGIS-X source list and the {\it getpsf} routine provided by \citet{laird09} was used to obtain their PSFs. To avoid contamination, the identified X-ray point sources in the mapped region were masked using the ellipse corresponding to the 95\% of the encircled energy radius \citep[EER; C15,][]{laird09}. Circles with radii that are $1.5$ times the semi-major axis of these ellipses are assigned zero counts and are placed on the maps with the contaminating point sources at the centre of these circles. The conservative circular masks used for each sources embed the entire PSF ellipse and hence the orientation angles of the PSFs do not play a significant role in our analysis.

C15 showed that the surface brightness profiles are statistically invariant if the used mask sizes  are 1 or 3 times the 95\% EER. A stacked source map is obtained by co-adding these individual source maps. A mask map is prepared accordingly. The pixels other than those falling within the circular masks around the point sources are assigned a value of 1 in the mask map. The mask maps are co-added to obtain a stacked mask map. The final average stacked image is obtained by dividing the stacked source map by the stacked mask map. All the gaps and singularities arising in the EAEC or the stacked maps are assigned null counts. Fig.\ 3 shows the final stacked maps of the AGN and the control sample respectively. The  original dimension of the stacked maps are of $300''\times 300''$, but only a $40''\times 40''$ square region around the centre has been shown in Fig.\ 3. Each pixel of the map corresponds to $0.5$ arcseconds.

\begin{figure*}[t]
\begin{center}
\begin{tabular}{c}
        \resizebox{8cm}{!}{\includegraphics{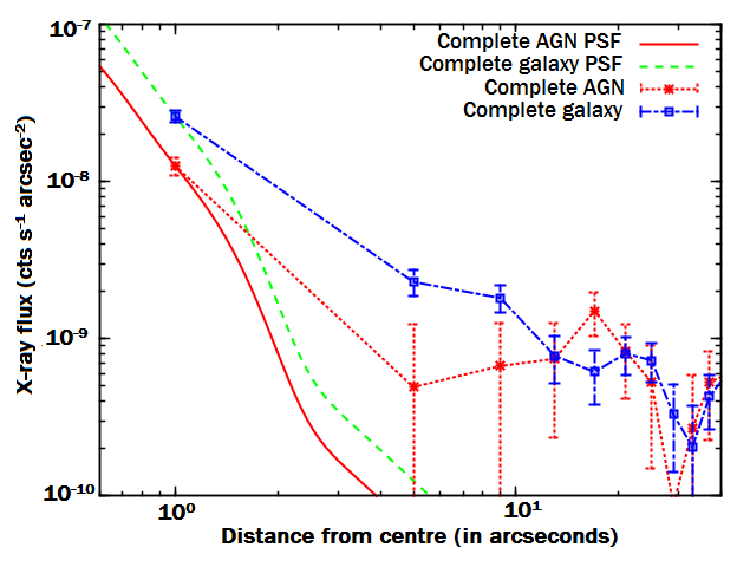}}
         \resizebox{8cm}{!}{\includegraphics{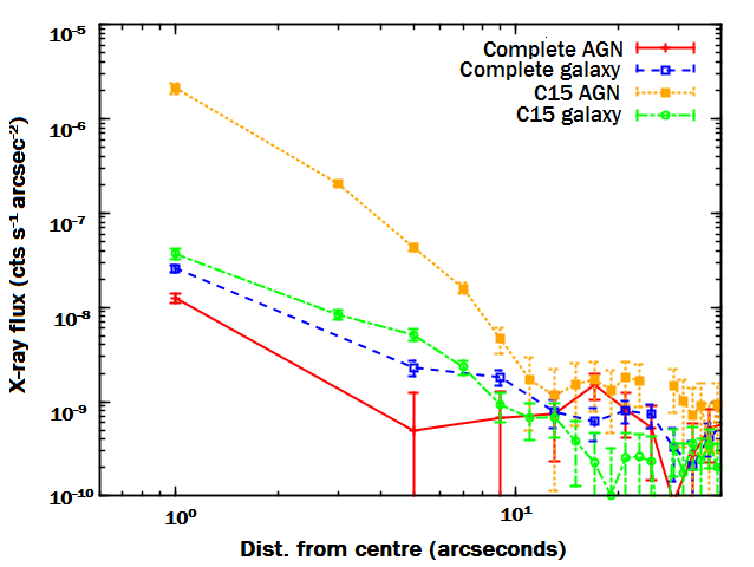}}  \\
          \resizebox{8cm}{!}{\includegraphics{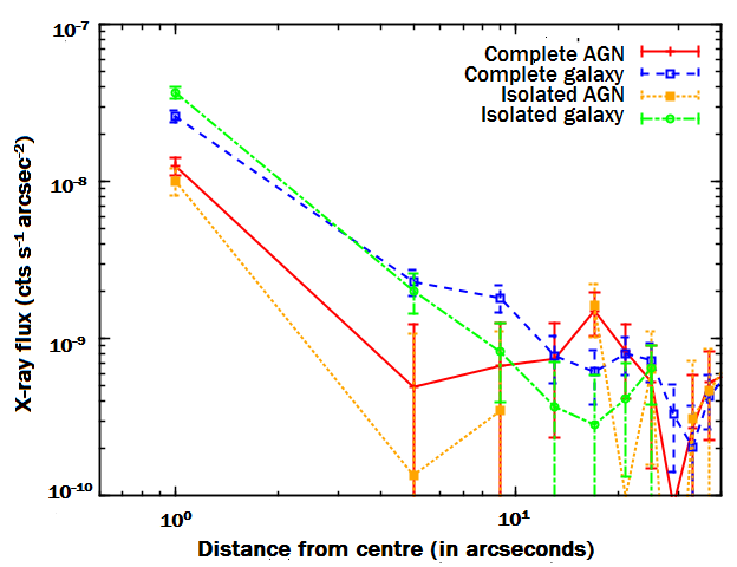}}  
          \resizebox{8cm}{!}{\includegraphics{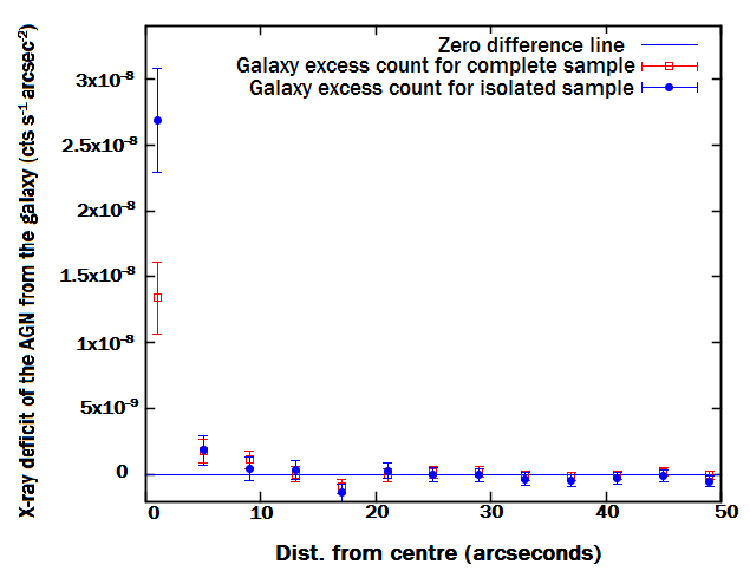}}
            
          \end{tabular}
 \caption{The {\bf top-left panel} shows the extended X-ray count plotted against distance from the centre of the complete sample stacked map with $ 0.3$\textless z\textless $0.6$ for both AGN and galaxies. The blue dot-dashed and the red dotted lines show the surface brightness profiles for the galaxy and the AGN stacked map respectively. The green dashed and the red solid lines show the PSF of the galaxy and the AGN stacked map respectively normalised by the maximum counts in each case. This plot corresponds to the complete sample of  AGN and Galaxy as described in \S 2 and Table 1. {\bf Top-right} panel shows the overplot of surface brightness profile of our complete sample with that obtained by C15 which used X-ray classified AGN for their analysis. The blue dashed line represents the profile for the current galaxy sample while the green dot-dashed line represents the galaxy profile for C15. The yellow dotted and the red solid lines represent the profiles of the C15 and current AGN samples respectively. C15 AGN sample (very high PSF) hugely dominates the C15 galaxy and as well as our AGN and galaxy complete samples. {\bf Bottom-left} panel shows the extended X-ray count plotted against distance from the centre of the isolated sources (see \S 2 for description of the sample) stacked map for both AGN and galaxies overplotted with the complete sample source surface brightness profile. The green dot-dashed and blue dashed lines represent the profiles for the isolated and complete galaxy samples respectively. The yellow dotted line and the red solid lines show the profiles for isolated and complete AGN. The {\bf bottom-right} panel  shows the galaxy count excess from the AGN counts calculated from the surface brightness profiles. For the complete sample the excess is $2.5$ $\sigma$ (red) and for isolated sources the galaxy excess is at a significance of $3$ $\sigma$ (blue). See \S 3 and \S 4 for more details.}
\end{center}
\label{Fig.4}
\end{figure*}

\section{Results}
The stacked maps of each of the two samples (AGN and control sample) have been used to extract the surface brightness profiles of AGN and galaxies. The average X-ray counts were extracted from the non-overlapping annuli each of width $4$ arcseconds in the stacked maps. The background counts were subtracted accordingly. Fig.\ 4 (top-left panel) shows the surface brightness profiles. The blue dot-dashed and the red dotted lines show the surface brightness profiles for the galaxy and the AGN respectively. The error-bars represent the standard errors at each annulus. The green dashed and the red solid lines show the average PSF of the galaxy and the AGN respectively normalised to the maximum counts in each case. The surface brightness profile of the galaxy clearly shows an excess as compared to that of the AGN.  

We note that the {\it Chandra} PSF depends on the position of the source in the detector and various other factors and hence would be different for each source. However since we use an average stacked map for the extraction of the surface brightness profile we compare our findings with that of the average {\it Chandra} PSF at soft X-ray energy following C15. A reasonable estimate of the PSF using the core and the wing is given by \citep{gaetz04}
\begin{eqnarray}
\mathnormal{f}(\theta)= \frac{A_0}{\bigg(1 +\left(\frac{\theta}{\theta_0}\right)^2\bigg)^{\gamma/2}} + A_1\exp\bigg[-4\ln(2)\left(\frac{\theta}{\theta_1}\right)^2\bigg] + \nonumber \\ 
A_2\exp\bigg[-4\ln(2)\left(\frac{\theta}{\theta_2}\right)^2\bigg]
\end{eqnarray}
The fit parameters are taken from C15 and \citet{gaetz04}. 

We stress that modeling the PSF profiles for our stacked maps is a crucial aspect of our analysis and it is important that the average analytic PSF profile closely mimics the actual PSF profile. Here we follow C15 to test for the differences between the average and the actual PSF profiles. In \S 2.3 we have described our method for obtaining the masks for the X-ray point sources. Using the same method we constructed a stacked PSF map of our sources. We note that the angular scale corresponding to 1.5 times the 95\% EER (the entire PSF profile) varies from $1''$ to about a maximum of $6''$ for all the AGN and galaxy sources used in the current work as well as C15. We note that the extent of the wings of the average PSF profiles are larger than the average EER. We further stress that the usage of large circular masks (see \S 2.3) would erase the difference in the axis angles of different source PSFs. We thus conclude that using the average PSF keeps our conclusion un-altered.

In the top right panel of Fig.\ 4 we compare our results with that of C15. The blue dashed line represents the profile for the current galaxy sample while the green dot-dashed line represents the galaxy profile for C15. The yellow dotted and the red solid lines represent the profiles of the C15 and current AGN samples respectively. The higher emission at the central region for the C15 AGN profile is coming from the bright X-ray emission from the accretion discs of AGN. It is evident and as has been noted in C15, the measured X-ray profile of their extended emission is contaminated by the PSF and the conclusions they draw in their paper are tentative. After correcting for PSF, C15 tend to find that the galaxy count was higher than the AGN within the error limits. With our new optically selected sources we observe that the excess in galaxy counts is significant compared to AGN. Our results thus favor the findings of C15. 

The stacked maps for the `isolated' samples are also subjected to the same procedure and the same analytic form of the average PSF is used to produce the X-ray surface brightness profiles of the isolated sources. The profiles are shown in the bottom-left panel of Fig.\ 4. We observe that the excess in galaxies over AGN increases when we exclude sources that are residing in group environments. 

Following C15 we obtain the difference profiles between galaxy and AGN. The error in the difference $\sigma$ is calculated as follows: $\sigma = \sqrt{ \sigma_A^2 + \sigma_G^2},$ where $\sigma_A$ is the standard error on the mean AGN count and $\sigma_G$ is the standard error on the mean of the galaxy at the same distance. The results are shown in the bottom right panel of Fig.\ 4. The positive points suggest an excess of the galaxy count over AGN and negative points suggest an excess of the AGN. The solid blue line represents no difference between the counts. We can see that the galaxy profile shows considerable excess as compared to the AGN.

We also compute the difference for the `isolated' samples using the same statistic described above. We note that the difference increases for the isolated sample when we exclude the AGN and galaxies residing in groups. The difference profile for the isolated sources has a significance of about 3 $\sigma$ and that of the complete sample shows a significance of about 2.5 $\sigma$. We also removed the group sources from C15 and performed the same analysis. We did not find any significant difference in the case of C15.

\section{Discussion}

Feedback from the central black hole and its role in galaxy evolution has been widely studied in the literature \citep [e.g.,][]{chatterjee15,hopkins16,cielo17,correa18, Pennyetal2018MNRAS}. The observational effects of AGN feedback has been estimated via many routes, the most recent of which involves the statistical detection of the signal through Sunyaev Zeldovich observations \citep[e.g.,][]{chatterjeeetal10, ruanetal15, verdieretal16, crichtonetal16, d&c17}. These large volume of theoretical and observational studies establish the importance and ubiquity of the role of AGN feedback in theories of structure formation. The direct effects of AGN feedback were seen in cluster and group environments through pioneering X-ray observations proving its global significance at all scales of structure formation. 

C15 tried to study the direct effect of AGN feedback on the soft X-ray gas in galaxies, extending the scope of the study to lower mass structures within the cosmic web. However, as mentioned before the study of C15 suffered from PSF contamination from X-ray point sources making their conclusions tentative. The current study aims at improving the conclusion drawn in C15 by replacing the X-ray selected AGN with the X-ray dim but optically bright sample of AGN. In this work, we detect a significant excess X-ray emission from the galaxy stacked maps as compared to the AGN stacked maps supporting the tentative conclusions of C15.

One of the key assumptions in theories of galaxy evolution is that galaxies reside in dark matter halos \citep[e.g.,][]{s&r98}. Thus the extended emission in halos is correlated with the mass of the dark matter halo that provides the gravitational potential well in which the galaxies reside \citep{w&f91,keres05,v&f11,correa18}. Studies show that even the stellar mass in galaxies is correlated with the host dark matter halo mass \citep[e.g.,][]{mosteretal10,shankar14,chiu16}. Observations suggest that pointed X-ray emission in galaxies too are correlated with stellar luminosity and stellar mass thereof 
\citep[e.g.,][]{formanetal85, t&f85, canizaresetal87, sarazinetal00,stricklandetal02, k&f03,  sivakoffetal03, fabbiano06, ranallietal08, mineoetal12}. It is believed that the relative contribution to the X-ray emission from the hot halo and the point sources might depend on galaxy luminosity and morphology (see \citealt {fabbiano06} and references therein).

Thus, to truly understand the X-ray emission in galaxies one requires a more detailed understanding of the X-ray source population. \citet{SC15} tried to characterise the X-ray source population by using galaxies from the DEEP-2 survey. Their results show that the X-ray emission in high redshift galaxies ($z \sim 0.6-0.7$) seem to be a combination of point sources as well as extended emission and a degeneracy existed between the two classes of X-ray emission. So one of the challenges in studies involving AGN feedback in high redshift galaxies, lies in lack of rigorous quantification of the sources emitting in X-rays. 

To alleviate this problem we have extracted the signal from two kinds of environment, namely group galaxies and isolated galaxies. It has been shown that group enviroment and dynamics have effects on the evolution of their member galaxies \citep[e.g.,][]{hou13,roberts17} and the IGM also is affected by the member galaxies \citep[e.g.,][]{heldson00,heldson03}. Hence we expect the galaxies belonging to the groups to be different from the isolated sources. The group galaxies are themselves a part of the larger gravitational well of the dark-matter halo. The diffuse X-ray due to this halo has a large contribution to the  X-ray of the member galaxies \citep{heldson01,memola09}. In our study we observe that while including the group galaxies in our stacking analysis provides an excess of the X-ray emission in normal over active galaxies, the significance of this excess increases when we exclude the group galaxies from our AGN and control galaxy samples (as shown in the bottom panels of Fig.\ 4). This is expected since the diffuse emission in groups in both the AGN and the control samples will tend to be identical and will smear out the differences in the average stacked signals. We removed the group belonging sources from C15 and did not find any difference in the surface brightness profiles from those obtained by C15. This is expected since the PSF emission is large in the case of C15. 

We calculate the bolometric luminosities of the optical AGN according to the methodology described in \citet{marconi04} using the B-mand magnitude and the redshifts. The bolometric luminosities vary between $10^{43}$-$10^{44}$ ergs $s^{-1}$. The accretion rates are calculated assuming the efficiency of conversion of accreted energy to radiated energy to be $10 \%$ \citep{s&s73}. These give accretion rates between $0.01 \Msun yr^{-1}$ to $0.17 \Msun yr^{-1}$. However we note that the value $10 \%$ is valid for radiatively efficient thin disc accretion, which is probably not the case in our AGN samples at large. Following the procedure described in \S 4 of C15 we obtain an estimate of feedback energy for our AGN sources. The feedback energy is found to be $2.1 \times {10^{39}}$ ergs $s^{-1}$ and $2.3 \times {10^{39}}$ ergs $s^{-1}$ respectively for the complete and the isolated AGN sources. Thus according to the current model the feedback energy is about $0.01$-$0.001 \%$ of the bolometric luminosity of the AGN. 

Now, if we assume the accretion rate energy to bolometric luminosity conversion efficiency to be $\eta$, the feedback fraction to be $\epsilon$, then the ratio of the feedback energy to that of the accretion energy will be $\eta\epsilon$. Observational studies of wind outflow from quasars find the kinetic luminosity of the outflow to be about $5-10 \%$  or in some cases $0.01-1 \%$ \citep{feruglio10,revalski18,crenshaw12,harrison14,husemann16} of the bolometric luminosities of these quasars. In the case of radio loud sources it has been observed that the kinetic energy of the jet is higher than the bolometric luminosity of the AGN.  Thus the parameter $\eta\epsilon$ in the above cases are higher than our estimates. 

Numerical simulations have revealed that jet couple with the ISM very efficiently \citep{dipanjan16,wagner12} and this efficiency is higher than radiative outflows \citep{cielo18}. Radiative feedback in cosmological simulations have assumed the feedback energy fractions ($\eta\epsilon$) to be typically $0.5 \%$ \citep[e.g.,][]{Dimatteoetal2005Natur, khandaietal15} of the accreted energy, consistent with our observed values. We stress that these simple estimates are representative and for more precise comparison we require understanding of accretion models in optically selected lower luminosity AGN.  

According to current studies, the role of AGN feedback in galaxy formation has been strong. Thus, different observational strategies are being exploited to asses the effect at different scales of structures in the Universe. Recently a direct meausure of AGN feedback through Sunyaev-Zeldovich observations with the Atacama Large Millimeter Array has been reported \citep{lacyetal19}. We like to stress that our technique to detect this effect at high redshifts, through X-ray observations is promising (with the limitations of the uncertainty in X-ray source populations). We thus propose to further carry out this study with samples of radio loud as well as obscure quasars to validate the evidence of AGN feedback using the X-ray staking technique. 

\section*{Acknowledgements}
SC acknowledges support from the Department of Science and Technology (Govt. of India) through the SERB Early Career Research grant and Presidency University through the Faculty Research and Professional Development Grant. SC is grateful to the Inter University Center for Astronomy and Astrophysics (IUCAA) for providing infra-structural and financial support along with local hospitality through the IUCAA-associateship program. SM is supported through the INSPIRE fellowship from the Department of Science and Technology, Govt. of India and the JBNSTS fellowship. SC and SM thank Arijit Sar for some help with the initial analysis of the work and an anonymous referee for suggesting several clarifications which helped in improving the draft. 

\nocite{*}

\bibliographystyle{apj}
\bibliography{SM_manuscript_revised}{}

\begin{thebibliography}{}
\expandafter\ifx\csname natexlab\endcsname\relax\def\natexlab#1{#1}\fi

\bibitem[{{Anderson} {et~al.}(2013){Anderson}, {Bregman}, \&
  {Dai}}]{anderson13}
{Anderson}, M.~E., {Bregman}, J.~N., \& {Dai}, X. 2013, \apj, 762, 106

\bibitem[{{Arnaud} \& {Evrard}(1999)}]{a&e99}
{Arnaud}, M., \& {Evrard}, A.~E. 1999, \mnras, 305, 631

\bibitem[{{Balbus} \& {Reynolds}(2010)}]{cr10}
{Balbus}, S.~A., \& {Reynolds}, C.~S. 2010, \apjl, 720, L97

\bibitem[{{Baldwin} {et~al.}(1981){Baldwin}, {Phillips}, \&
  {Terlevich}}]{bpt81}
{Baldwin}, J.~A., {Phillips}, M.~M., \& {Terlevich}, R. 1981, \pasp, 93, 5

\bibitem[{{Battaglia} {et~al.}(2010){Battaglia}, {Bond}, {Pfrommer}, {Sievers},
  \& {Sijacki}}]{battagliaetal10}
{Battaglia}, N., {Bond}, J.~R., {Pfrommer}, C., {Sievers}, J.~L., \& {Sijacki},
  D. 2010, \apj, 725, 91

\bibitem[{{Beifiori} {et~al.}(2014){Beifiori}, {Thomas}, {Maraston}, {Steele},
  {Masters}, {Pforr}, {Saglia}, {Bender}, {Tojeiro}, {Chen}, {Bolton},
  {Brownstein}, {Johansson}, {Leauthaud}, {Nichol}, {Schneider}, {Senger},
  {Skibba}, {Wake}, {Pan}, {Snedden}, {Bizyaev}, {Brewington}, {Malanushenko},
  {Malanushenko}, {Oravetz}, {Simmons}, {Shelden}, \& {Ebelke}}]{beifori14}
{Beifiori}, A., {Thomas}, D., {Maraston}, C., {et~al.} 2014, \apj, 789, 92

\bibitem[{{Bhattacharya} {et~al.}(2008){Bhattacharya}, {Di Matteo}, \&
  {Kosowsky}}]{bhattacharyaetal08}
{Bhattacharya}, S., {Di Matteo}, T., \& {Kosowsky}, A. 2008, \mnras, 389, 34

\bibitem[{{Bogdanovi{\'c}} {et~al.}(2009){Bogdanovi{\'c}}, {Reynolds},
  {Balbus}, \& {Parrish}}]{cr09}
{Bogdanovi{\'c}}, T., {Reynolds}, C.~S., {Balbus}, S.~A., \& {Parrish}, I.~J.
  2009, \apj, 704, 211

\bibitem[{{Canizares} {et~al.}(1987){Canizares}, {Fabbiano}, \&
  {Trinchieri}}]{canizaresetal87}
{Canizares}, C.~R., {Fabbiano}, G., \& {Trinchieri}, G. 1987, \apj, 312, 503

\bibitem[{Cavagnolo {et~al.}(2011)Cavagnolo, McNamara, Wise, Nulsen, Brüggen,
  Gitti, \& Rafferty}]{birzan04}
Cavagnolo, K.~W., McNamara, B.~R., Wise, M.~W., {et~al.} 2011, The
  Astrophysical Journal, 732, 71

\bibitem[{{Chatterjee} {et~al.}(2008){Chatterjee}, {Di Matteo}, {Kosowsky}, \&
  {Pelupessy}}]{chatterjeeetal08}
{Chatterjee}, S., {Di Matteo}, T., {Kosowsky}, A., \& {Pelupessy}, I. 2008,
  \mnras, 390, 535

\bibitem[{{Chatterjee} {et~al.}(2010){Chatterjee}, {Ho}, {Newman}, \&
  {Kosowsky}}]{chatterjeeetal10}
{Chatterjee}, S., {Ho}, S., {Newman}, J.~A., \& {Kosowsky}, A. 2010, \apj, 720,
  299

\bibitem[{{Chatterjee} \& {Kosowsky}(2007)}]{c&k07}
{Chatterjee}, S., \& {Kosowsky}, A. 2007, \apjl, 661, L113

\bibitem[{{Chatterjee} {et~al.}(2015{\natexlab{a}}){Chatterjee}, {Newman},
  {Jeltema}, {Myers}, {Aird}, {Bundy}, {Conselice}, {Cooper}, {Laird},
  {Nandra}, \& {Willmer}}]{SC15}
{Chatterjee}, S., {Newman}, J.~A., {Jeltema}, T., {et~al.} 2015{\natexlab{a}},
  \apj, 806, 136

\bibitem[{{Chatterjee} {et~al.}(2015{\natexlab{b}}){Chatterjee}, {Newman},
  {Jeltema}, {Myers}, {Aird}, {Coil}, {Cooper}, {Finoguenov}, {Laird},
  {Montero-Dorta}, {Nandra}, {Willmer}, \& {Yan}}]{chatterjee15}
---. 2015{\natexlab{b}}, \pasp, 127, 716

\bibitem[{{Chaudhuri} {et~al.}(2013){Chaudhuri}, {Majumdar}, \&
  {Nath}}]{Chaudhurietal2013ApJ}
{Chaudhuri}, A., {Majumdar}, S., \& {Nath}, B.~B. 2013, \apj, 776, 84

\bibitem[{{Chiu} {et~al.}(2016){Chiu}, {Saro}, {Mohr}, {Desai}, {Bocquet},
  {Capasso}, {Gangkofner}, {Gupta}, \& {Liu}}]{chiu16}
{Chiu}, I., {Saro}, A., {Mohr}, J., {et~al.} 2016, \mnras, 458, 379

\bibitem[{{Choi} {et~al.}(2013){Choi}, {Naab}, {Ostriker}, {Johansson}, \&
  {Moster}}]{choietal13}
{Choi}, E., {Naab}, T., {Ostriker}, J.~P., {Johansson}, P.~H., \& {Moster},
  B.~P. 2013, ArXiv e-prints, arXiv:1308.3719

\bibitem[{{Choi} {et~al.}(2004){Choi}, {Reynolds}, {Heinz}, {Rosenberg},
  {Perlman}, \& {Yang}}]{choi04}
{Choi}, Y.-Y., {Reynolds}, C.~S., {Heinz}, S., {et~al.} 2004, \apj, 606, 185

\bibitem[{{Cielo} {et~al.}(2017){Cielo}, {Bieri}, {Volonteri}, {Wagner}, \&
  {Dubois}}]{cielo17}
{Cielo}, S., {Bieri}, R., {Volonteri}, M., {Wagner}, A., \& {Dubois}, Y. 2017,
  ArXiv e-prints, arXiv:1712.03955

\bibitem[{{Cielo} {et~al.}(2018){Cielo}, {Bieri}, {Volonteri}, {Wagner}, \&
  {Dubois}}]{cielo18}
{Cielo}, S., {Bieri}, R., {Volonteri}, M., {Wagner}, A.~Y., \& {Dubois}, Y.
  2018, \mnras, 477, 1336

\bibitem[{{Ciotti} \& {Ostriker}(2001)}]{ciotti01}
{Ciotti}, L., \& {Ostriker}, J.~P. 2001, \apj, 551, 131

\bibitem[{{Correa} {et~al.}(2018){Correa}, {Schaye}, {Wyithe}, {Duffy},
  {Theuns}, {Crain}, \& {Bower}}]{correa18}
{Correa}, C.~A., {Schaye}, J., {Wyithe}, J.~S.~B., {et~al.} 2018, \mnras, 473,
  538

\bibitem[{{Cox} {et~al.}(2006){Cox}, {Dutta}, {Di Matteo}, {Hernquist},
  {Hopkins}, {Robertson}, \& {Springel}}]{Coxetal2006ApJ}
{Cox}, T.~J., {Dutta}, S.~N., {Di Matteo}, T., {et~al.} 2006, \apj, 650, 791

\bibitem[{{Crenshaw} \& {Kraemer}(2012)}]{crenshaw12}
{Crenshaw}, D.~M., \& {Kraemer}, S.~B. 2012, \apj, 753, 75

\bibitem[{{Crichton} {et~al.}(2016){Crichton}, {Gralla}, {Hall}, {Marriage},
  {Zakamska}, {Battaglia}, {Bond}, {Devlin}, {Hill}, {Hilton}, {Hincks},
  {Huffenberger}, {Hughes}, {Kosowsky}, {Moodley}, {Niemack}, {Page},
  {Partridge}, {Sievers}, {Sif{\'o}n}, {Staggs}, {Viero}, \&
  {Wollack}}]{crichtonetal16}
{Crichton}, D., {Gralla}, M.~B., {Hall}, K., {et~al.} 2016, \mnras, 458, 1478

\bibitem[{{Croton} {et~al.}(2006){Croton}, {Springel}, {White}, {De Lucia},
  {Frenk}, {Gao}, {Jenkins}, {Kauffmann}, {Navarro}, \&
  {Yoshida}}]{crotonetal06}
{Croton}, D.~J., {Springel}, V., {White}, S.~D.~M., {et~al.} 2006, \mnras, 365,
  11

\bibitem[{{Davis} {et~al.}(2003){Davis}, {Faber}, {Newman}, {Phillips},
  {Ellis}, {Steidel}, {Conselice}, {Coil}, {Finkbeiner}, {Koo}, {Guhathakurta},
  {Weiner}, {Schiavon}, {Willmer}, {Kaiser}, {Luppino}, {Wirth}, {Connolly},
  {Eisenhardt}, {Cooper}, \& {Gerke}}]{davis03}
{Davis}, M., {Faber}, S.~M., {Newman}, J., {et~al.} 2003, in \procspie, Vol.
  4834, Discoveries and Research Prospects from 6- to 10-Meter-Class Telescopes
  II, ed. P.~{Guhathakurta}, 161--172

\bibitem[{{Davis} {et~al.}(2007){Davis}, {Guhathakurta}, {Konidaris}, {Newman},
  {Ashby}, {Biggs}, {Barmby}, {Bundy}, {Chapman}, {Coil}, {Conselice},
  {Cooper}, {Croton}, {Eisenhardt}, {Ellis}, {Faber}, {Fang}, {Fazio},
  {Georgakakis}, {Gerke}, {Goss}, {Gwyn}, {Harker}, {Hopkins}, {Huang},
  {Ivison}, {Kassin}, {Kirby}, {Koekemoer}, {Koo}, {Laird}, {Le Floc'h}, {Lin},
  {Lotz}, {Marshall}, {Martin}, {Metevier}, {Moustakas}, {Nandra}, {Noeske},
  {Papovich}, {Phillips}, {Rich}, {Rieke}, {Rigopoulou}, {Salim},
  {Schiminovich}, {Simard}, {Smail}, {Small}, {Weiner}, {Willmer}, {Willner},
  {Wilson}, {Wright}, \& {Yan}}]{davisetal07}
{Davis}, M., {Guhathakurta}, P., {Konidaris}, N.~P., {et~al.} 2007, \apjl, 660,
  L1

\bibitem[{{Di Matteo} {et~al.}(2005){Di Matteo}, {Springel}, \&
  {Hernquist}}]{Dimatteoetal2005Natur}
{Di Matteo}, T., {Springel}, V., \& {Hernquist}, L. 2005, \nat, 433, 604

\bibitem[{Doria {et~al.}(2012)Doria, Gitti, Ettori, Brighenti, Nulsen, \&
  McNamara}]{doria12}
Doria, A., Gitti, M., Ettori, S., {et~al.} 2012, The Astrophysical Journal,
  753, 47

\bibitem[{{Dunn} \& {Fabian}(2006{\natexlab{a}})}]{d&f06}
{Dunn}, R.~J.~H., \& {Fabian}, A.~C. 2006{\natexlab{a}}, \mnras, 373, 959

\bibitem[{{Dunn} \& {Fabian}(2006{\natexlab{b}})}]{dunnfabian06}
---. 2006{\natexlab{b}}, \mnras, 373, 959

\bibitem[{{Dutta Chowdhury} \& {Chatterjee}(2017)}]{d&c17}
{Dutta Chowdhury}, D., \& {Chatterjee}, S. 2017, \apj, 839, 34

\bibitem[{{Fabbiano}(2006)}]{fabbiano06}
{Fabbiano}, G. 2006, \araa, 44, 323

\bibitem[{{Fabian} {et~al.}(2005){Fabian}, {Reynolds}, {Taylor}, \&
  {Dunn}}]{fab05}
{Fabian}, A.~C., {Reynolds}, C.~S., {Taylor}, G.~B., \& {Dunn}, R.~J.~H. 2005,
  \mnras, 363, 891

\bibitem[{{Fertig} {et~al.}(2014){Fertig}, {Rosenberg}, {Patton}, \&
  {Ellison}}]{fertig14}
{Fertig}, D., {Rosenberg}, J.~L., {Patton}, D.~R., \& {Ellison}, S.~L. 2014, in
  American Astronomical Society Meeting Abstracts, Vol. 223, American
  Astronomical Society Meeting Abstracts \#223, 246.15

\bibitem[{{Feruglio} {et~al.}(2010){Feruglio}, {Maiolino}, {Piconcelli},
  {Menci}, {Aussel}, {Lamastra}, \& {Fiore}}]{feruglio10}
{Feruglio}, C., {Maiolino}, R., {Piconcelli}, E., {et~al.} 2010, \aap, 518,
  L155

\bibitem[{{Forman} {et~al.}(1985){Forman}, {Jones}, \& {Tucker}}]{formanetal85}
{Forman}, W., {Jones}, C., \& {Tucker}, W. 1985, \apj, 293, 102

\bibitem[{{Gaetz} {et~al.}(2004){Gaetz}, {Edgar}, {Jerius}, {Zhao}, \&
  {Smith}}]{gaetz04}
{Gaetz}, T.~J., {Edgar}, R.~J., {Jerius}, D.~H., {Zhao}, P., \& {Smith}, R.~K.
  2004, in \procspie, Vol. 5165, X-Ray and Gamma-Ray Instrumentation for
  Astronomy XIII, ed. K.~A. {Flanagan} \& O.~H.~W. {Siegmund}, 411--422

\bibitem[{{Gaspari} {et~al.}(2012){Gaspari}, {Brighenti}, \&
  {Temi}}]{gasparietal12}
{Gaspari}, M., {Brighenti}, F., \& {Temi}, P. 2012, \mnras, 424, 190

\bibitem[{{Gebhardt} {et~al.}(2000){Gebhardt}, {Bender}, {Bower}, {Dressler},
  {Faber}, {Filippenko}, {Green}, {Grillmair}, {Ho}, {Kormendy}, {Lauer},
  {Magorrian}, {Pinkney}, {Richstone}, \& {Tremaine}}]{gebhardtetal00}
{Gebhardt}, K., {Bender}, R., {Bower}, G., {et~al.} 2000, \apjl, 539, L13

\bibitem[{{Gerke} {et~al.}(2012){Gerke}, {Newman}, {Davis}, {Coil}, {Cooper},
  {Dutton}, {Faber}, {Guhathakurta}, {Konidaris}, {Koo}, {Lin}, {Noeske},
  {Phillips}, {Rosario}, {Weiner}, {Willmer}, \& {Yan}}]{gerke12}
{Gerke}, B.~F., {Newman}, J.~A., {Davis}, M., {et~al.} 2012, \apj, 751, 50

\bibitem[{Giacintucci {et~al.}(2011)Giacintucci, O’Sullivan, Vrtilek, David,
  Raychaudhury, Venturi, Athreya, Clarke, Murgia, Mazzotta, Gitti, Ponman,
  Ishwara-Chandra, Jones, \& Forman}]{src11}
Giacintucci, S., O’Sullivan, E., Vrtilek, J., {et~al.} 2011, The
  Astrophysical Journal, 732, 95

\bibitem[{{Giodini} {et~al.}(2010){Giodini}, {Smol{\v c}i{\'c}}, {Finoguenov},
  {Boehringer}, {B{\^i}rzan}, {Zamorani}, {Oklop{\v c}i{\'c}}, {Pierini},
  {Pratt}, {Schinnerer}, {Massey}, {Koekemoer}, {Salvato}, {Sanders},
  {Kartaltepe}, \& {Thompson}}]{giodini10}
{Giodini}, S., {Smol{\v c}i{\'c}}, V., {Finoguenov}, A., {et~al.} 2010, \apj,
  714, 218

\bibitem[{{Gitti} {et~al.}(2012){Gitti}, {Brighenti}, \&
  {McNamara}}]{gittietal12}
{Gitti}, M., {Brighenti}, F., \& {McNamara}, B.~R. 2012, Advances in Astronomy,
  2012, arXiv:1109.3334

\bibitem[{{Goulding} {et~al.}(2011){Goulding}, {Alexander}, {Mullaney},
  {Gelbord}, {Hickox}, {Ward}, \& {Watson}}]{goulding11}
{Goulding}, A.~D., {Alexander}, D.~M., {Mullaney}, J.~R., {et~al.} 2011,
  \mnras, 411, 1231

\bibitem[{{Harrison} {et~al.}(2014){Harrison}, {Alexander}, {Mullaney}, \&
  {Swinbank}}]{harrison14}
{Harrison}, C.~M., {Alexander}, D.~M., {Mullaney}, J.~R., \& {Swinbank}, A.~M.
  2014, \mnras, 441, 3306

\bibitem[{{Heinz} {et~al.}(2002){Heinz}, {Choi}, {Reynolds}, \&
  {Begelman}}]{cr02}
{Heinz}, S., {Choi}, Y.-Y., {Reynolds}, C.~S., \& {Begelman}, M.~C. 2002,
  \apjl, 569, L79

\bibitem[{{Heinz} {et~al.}(1998){Heinz}, {Reynolds}, \& {Begelman}}]{cr98}
{Heinz}, S., {Reynolds}, C.~S., \& {Begelman}, M.~C. 1998, \apj, 501, 126

\bibitem[{{Helsdon} \& {Ponman}(2000)}]{heldson00}
{Helsdon}, S.~F., \& {Ponman}, T.~J. 2000, \mnras, 315, 356

\bibitem[{{Helsdon} \& {Ponman}(2003)}]{heldson03}
---. 2003, \mnras, 340, 485

\bibitem[{{Helsdon} {et~al.}(2001){Helsdon}, {Ponman}, {O'Sullivan}, \&
  {Forbes}}]{heldson01}
{Helsdon}, S.~F., {Ponman}, T.~J., {O'Sullivan}, E., \& {Forbes}, D.~A. 2001,
  \mnras, 325, 693

\bibitem[{{Hopkins} {et~al.}(2006){Hopkins}, {Robertson}, {Krause},
  {Hernquist}, \& {Cox}}]{hopkinsetal06}
{Hopkins}, P.~F., {Robertson}, B., {Krause}, E., {Hernquist}, L., \& {Cox},
  T.~J. 2006, \apj, 652, 107

\bibitem[{{Hopkins} {et~al.}(2016){Hopkins}, {Torrey}, {Faucher-Gigu{\`e}re},
  {Quataert}, \& {Murray}}]{hopkins16}
{Hopkins}, P.~F., {Torrey}, P., {Faucher-Gigu{\`e}re}, C.-A., {Quataert}, E.,
  \& {Murray}, N. 2016, \mnras, 458, 816

\bibitem[{{Hou} {et~al.}(2013){Hou}, {Parker}, {Balogh}, {McGee}, {Wilman},
  {Connelly}, {Harris}, {Mok}, {Mulchaey}, {Bower}, \& {Finoguenov}}]{hou13}
{Hou}, A., {Parker}, L.~C., {Balogh}, M.~L., {et~al.} 2013, \mnras, 435, 1715

\bibitem[{{Husemann} {et~al.}(2016){Husemann}, {Scharw{\"a}chter}, {Bennert},
  {Mainieri}, {Woo}, \& {Kakkad}}]{husemann16}
{Husemann}, B., {Scharw{\"a}chter}, J., {Bennert}, V.~N., {et~al.} 2016, \aap,
  594, A44

\bibitem[{{Kaiser} \& {Binney}(2003)}]{Kaiseretal2003MNRAS}
{Kaiser}, C.~R., \& {Binney}, J. 2003, \mnras, 338, 837

\bibitem[{{Kauffmann} {et~al.}(2003){Kauffmann}, {Heckman}, {Tremonti},
  {Brinchmann}, {Charlot}, {White}, {Ridgway}, {Brinkmann}, {Fukugita}, {Hall},
  {Ivezi{\'c}}, {Richards}, \& {Schneider}}]{kauffman03}
{Kauffmann}, G., {Heckman}, T.~M., {Tremonti}, C., {et~al.} 2003, \mnras, 346,
  1055

\bibitem[{{Kere{\v s}} {et~al.}(2005){Kere{\v s}}, {Katz}, {Weinberg}, \&
  {Dav{\'e}}}]{keres05}
{Kere{\v s}}, D., {Katz}, N., {Weinberg}, D.~H., \& {Dav{\'e}}, R. 2005,
  \mnras, 363, 2

\bibitem[{{Kewley} {et~al.}(2001){Kewley}, {Dopita}, {Sutherland}, {Heisler},
  \& {Trevena}}]{kewley01}
{Kewley}, L.~J., {Dopita}, M.~A., {Sutherland}, R.~S., {Heisler}, C.~A., \&
  {Trevena}, J. 2001, \apj, 556, 121

\bibitem[{{Khandai} {et~al.}(2015){Khandai}, {Di Matteo}, {Croft}, {Wilkins},
  {Feng}, {Tucker}, {DeGraf}, \& {Liu}}]{khandaietal15}
{Khandai}, N., {Di Matteo}, T., {Croft}, R., {et~al.} 2015, \mnras, 450, 1349

\bibitem[{{Kim} \& {Fabbiano}(2003)}]{k&f03}
{Kim}, D.-W., \& {Fabbiano}, G. 2003, \apj, 586, 826

\bibitem[{{Komossa} \& {B{\"o}hringer}(1999)}]{komossa99}
{Komossa}, S., \& {B{\"o}hringer}, H. 1999, in Diffuse Thermal and Relativistic
  Plasma in Galaxy Clusters, ed. H.~{Boehringer}, L.~{Feretti}, \&
  P.~{Schuecker}, 167

\bibitem[{{Kunz} {et~al.}(2012){Kunz}, {Bogdanovi{\'c}}, {Reynolds}, \&
  {Stone}}]{cr12}
{Kunz}, M.~W., {Bogdanovi{\'c}}, T., {Reynolds}, C.~S., \& {Stone}, J.~M. 2012,
  \apj, 754, 122

\bibitem[{{Kurosawa} {et~al.}(2009){Kurosawa}, {Proga}, \&
  {Nagamine}}]{kurosawa09}
{Kurosawa}, R., {Proga}, D., \& {Nagamine}, K. 2009, \apj, 707, 823

\bibitem[{{Lacy} {et~al.}(2019){Lacy}, {Mason}, {Sarazin}, {Chatterjee},
  {Nyland}, {Kimball}, {Rocha}, {Rowe}, \& {Surace}}]{lacyetal19}
{Lacy}, M., {Mason}, B., {Sarazin}, C., {et~al.} 2019, \mnras, 483, L22

\bibitem[{{Laird} {et~al.}(2009){Laird}, {Nandra}, {Georgakakis}, {Aird},
  {Barmby}, {Conselice}, {Coil}, {Davis}, {Faber}, {Fazio}, {Guhathakurta},
  {Koo}, {Sarajedini}, \& {Willmer}}]{laird09}
{Laird}, E.~S., {Nandra}, K., {Georgakakis}, A., {et~al.} 2009, \apjs, 180, 102

\bibitem[{{Lapi} {et~al.}(2014){Lapi}, {Raimundo}, {Aversa}, {Cai}, {Negrello},
  {Celotti}, {De Zotti}, \& {Danese}}]{lapi14}
{Lapi}, A., {Raimundo}, S., {Aversa}, R., {et~al.} 2014, \apj, 782, 69

\bibitem[{{Marconi} {et~al.}(2004){Marconi}, {Risaliti}, {Gilli}, {Hunt},
  {Maiolino}, \& {Salvati}}]{marconi04}
{Marconi}, A., {Risaliti}, G., {Gilli}, R., {et~al.} 2004, \mnras, 351, 169

\bibitem[{{McNamara} \& {Nulsen}(2007{\natexlab{a}})}]{m&n07}
{McNamara}, B.~R., \& {Nulsen}, P.~E.~J. 2007{\natexlab{a}}, \araa, 45, 117

\bibitem[{{McNamara} \& {Nulsen}(2007{\natexlab{b}})}]{mcnamaranulsen07}
---. 2007{\natexlab{b}}, \araa, 45, 117

\bibitem[{McNamara {et~al.}(2016)McNamara, Russell, Nulsen, Hogan, Fabian,
  Pulido, \& Edge}]{birzan}
McNamara, B.~R., Russell, H.~R., Nulsen, P. E.~J., {et~al.} 2016, The
  Astrophysical Journal, 830, 79

\bibitem[{{Memola} {et~al.}(2009){Memola}, {Trinchieri}, {Wolter}, {Focardi},
  \& {Kelm}}]{memola09}
{Memola}, E., {Trinchieri}, G., {Wolter}, A., {Focardi}, P., \& {Kelm}, B.
  2009, \aap, 497, 359

\bibitem[{{Merritt} \& {Ferrarese}(2001{\natexlab{a}})}]{m&f01}
{Merritt}, D., \& {Ferrarese}, L. 2001{\natexlab{a}}, \apj, 547, 140

\bibitem[{{Merritt} \& {Ferrarese}(2001{\natexlab{b}})}]{merritt01}
---. 2001{\natexlab{b}}, \apj, 547, 140

\bibitem[{{Mineo} {et~al.}(2012){Mineo}, {Gilfanov}, \&
  {Sunyaev}}]{mineoetal12}
{Mineo}, S., {Gilfanov}, M., \& {Sunyaev}, R. 2012, \mnras, 426, 1870

\bibitem[{{Moster} {et~al.}(2010){Moster}, {Somerville}, {Maulbetsch}, {van den
  Bosch}, {Macci{\`o}}, {Naab}, \& {Oser}}]{mosteretal10}
{Moster}, B.~P., {Somerville}, R.~S., {Maulbetsch}, C., {et~al.} 2010, \apj,
  710, 903

\bibitem[{{Mukherjee} {et~al.}(2016){Mukherjee}, {Bicknell}, {Sutherland}, \&
  {Wagner}}]{dipanjan16}
{Mukherjee}, D., {Bicknell}, G.~V., {Sutherland}, R., \& {Wagner}, A. 2016,
  \mnras, 461, 967

\bibitem[{{Nath} \& {Roychowdhury}(2002)}]{n&r02}
{Nath}, B.~B., \& {Roychowdhury}, S. 2002, \mnras, 333, 145

\bibitem[{{Newman} {et~al.}(2013){Newman}, {Cooper}, {Davis}, {Faber}, {Coil},
  {Guhathakurta}, {Koo}, {Phillips}, {Conroy}, {Dutton}, {Finkbeiner}, {Gerke},
  {Rosario}, {Weiner}, {Willmer}, {Yan}, {Harker}, {Kassin}, {Konidaris},
  {Lai}, {Madgwick}, {Noeske}, {Wirth}, {Connolly}, {Kaiser}, {Kirby},
  {Lemaux}, {Lin}, {Lotz}, {Luppino}, {Marinoni}, {Matthews}, {Metevier}, \&
  {Schiavon}}]{newman13}
{Newman}, J.~A., {Cooper}, M.~C., {Davis}, M., {et~al.} 2013, \apjs, 208, 5

\bibitem[{Nulsen {et~al.}(2005)Nulsen, Hambrick, McNamara, Rafferty, Birzan,
  Wise, \& David}]{nulsen05}
Nulsen, P. E.~J., Hambrick, D.~C., McNamara, B.~R., {et~al.} 2005, The
  Astrophysical Journal Letters, 625, L9

\bibitem[{{Nulsen} {et~al.}(2005){Nulsen}, {Hambrick}, {McNamara}, {Rafferty},
  {Birzan}, {Wise}, \& {David}}]{nulsenetal05}
{Nulsen}, P.~E.~J., {Hambrick}, D.~C., {McNamara}, B.~R., {et~al.} 2005, \apjl,
  625, L9

\bibitem[{{O'Sullivan} {et~al.}(2011){O'Sullivan}, {Giacintucci}, {David},
  {Gitti}, {Vrtilek}, {Raychaudhury}, \& {Ponman}}]{david11}
{O'Sullivan}, E., {Giacintucci}, S., {David}, L.~P., {et~al.} 2011, \apj, 735,
  11

\bibitem[{{Pellegrini} {et~al.}(2012){Pellegrini}, {Ciotti}, \&
  {Ostriker}}]{pelligrinietal12}
{Pellegrini}, S., {Ciotti}, L., \& {Ostriker}, J.~P. 2012, \apj, 744, 21

\bibitem[{{Penny} {et~al.}(2018){Penny}, {Masters}, {Smethurst}, {Nichol},
  {Krawczyk}, {Bizyaev}, {Greene}, {Liu}, {Marinelli}, {Rembold}, {Riffel},
  {Ilha}, {Wylezalek}, {Andrews}, {Bundy}, {Drory}, {Oravetz}, \&
  {Pan}}]{Pennyetal2018MNRAS}
{Penny}, S.~J., {Masters}, K.~L., {Smethurst}, R., {et~al.} 2018, \mnras, 476,
  979

\bibitem[{{Peterson} \& {Fabian}(2006)}]{p&f06}
{Peterson}, J.~R., \& {Fabian}, A.~C. 2006, \physrep, 427, 1

\bibitem[{{Puchwein} {et~al.}(2010){Puchwein}, {Springel}, {Sijacki}, \&
  {Dolag}}]{puchweinetal10}
{Puchwein}, E., {Springel}, V., {Sijacki}, D., \& {Dolag}, K. 2010, \mnras,
  406, 936

\bibitem[{{Ranalli} {et~al.}(2008){Ranalli}, {Comastri}, {Origlia}, \&
  {Maiolino}}]{ranallietal08}
{Ranalli}, P., {Comastri}, A., {Origlia}, L., \& {Maiolino}, R. 2008, \mnras,
  386, 1464

\bibitem[{{Ranalli} {et~al.}(2003){Ranalli}, {Comastri}, \&
  {Setti}}]{ranalli03}
{Ranalli}, P., {Comastri}, A., \& {Setti}, G. 2003, \aap, 399, 39

\bibitem[{{Randall} {et~al.}(2014){Randall}, {Nulsen}, {Clarke}, {Forman},
  {Jones}, {Kraft}, \& {Blanton}}]{randall14}
{Randall}, S.~W., {Nulsen}, P., {Clarke}, T.~E., {et~al.} 2014, in American
  Astronomical Society Meeting Abstracts, Vol. 224, American Astronomical
  Society Meeting Abstracts \#224, 222.05

\bibitem[{{Randall} {et~al.}(2015){Randall}, {Nulsen}, {Jones}, {Forman},
  {Clarke}, \& {Blanton}}]{randall15}
{Randall}, S.~W., {Nulsen}, P.~E.~J., {Jones}, C., {et~al.} 2015, in IAU
  Symposium, Vol. 313, Extragalactic Jets from Every Angle, ed. F.~{Massaro},
  C.~C. {Cheung}, E.~{Lopez}, \& A.~{Siemiginowska}, 277--282

\bibitem[{Randall {et~al.}(2011)Randall, Forman, Giacintucci, Nulsen, Sun,
  Jones, Churazov, David, Kraft, Donahue, Blanton, Simionescu, \&
  Werner}]{Randall11}
Randall, S.~W., Forman, W.~R., Giacintucci, S., {et~al.} 2011, The
  Astrophysical Journal, 726, 86

\bibitem[{{Revalski} {et~al.}(2018){Revalski}, {Crenshaw}, {Kraemer},
  {Fischer}, {Schmitt}, \& {Machuca}}]{revalski18}
{Revalski}, M., {Crenshaw}, D.~M., {Kraemer}, S.~B., {et~al.} 2018, \apj, 856,
  46

\bibitem[{{Reynolds} {et~al.}(2005{\natexlab{a}}){Reynolds}, {Brenneman}, \&
  {Stocke}}]{cr105}
{Reynolds}, C.~S., {Brenneman}, L.~W., \& {Stocke}, J.~T. 2005{\natexlab{a}},
  \mnras, 357, 381

\bibitem[{{Reynolds} \& {Fabian}(1995)}]{cr95}
{Reynolds}, C.~S., \& {Fabian}, A.~C. 1995, \mnras, 273, 1167

\bibitem[{{Reynolds} {et~al.}(2001){Reynolds}, {Heinz}, \& {Begelman}}]{cr01}
{Reynolds}, C.~S., {Heinz}, S., \& {Begelman}, M.~C. 2001, \apjl, 549, L179

\bibitem[{{Reynolds} {et~al.}(2002){Reynolds}, {Heinz}, \& {Begelman}}]{cr102}
---. 2002, \mnras, 332, 271

\bibitem[{{Reynolds} {et~al.}(2005{\natexlab{b}}){Reynolds}, {McKernan},
  {Fabian}, {Stone}, \& {Vernaleo}}]{cr05}
{Reynolds}, C.~S., {McKernan}, B., {Fabian}, A.~C., {Stone}, J.~M., \&
  {Vernaleo}, J.~C. 2005{\natexlab{b}}, \mnras, 357, 242

\bibitem[{{Roberg-Clark} {et~al.}(2016){Roberg-Clark}, {Drake}, {Reynolds}, \&
  {Swisdak}}]{clark16}
{Roberg-Clark}, G.~T., {Drake}, J.~F., {Reynolds}, C.~S., \& {Swisdak}, M.
  2016, \apjl, 830, L9

\bibitem[{{Roberts} \& {Parker}(2017)}]{roberts17}
{Roberts}, I.~D., \& {Parker}, L.~C. 2017, in Galaxy Evolution Across Time,
  Proceedings of a conference held 12-16 June, 2017 in Paris. Online at <A
  href=``https://galaxiesinparis.sciencesconf.org/''>
  https://galaxiesinparis.sciencesconf.org/</A>, id. 12, 12

\bibitem[{{Ruan} {et~al.}(2015){Ruan}, {McQuinn}, \& {Anderson}}]{ruanetal15}
{Ruan}, J.~J., {McQuinn}, M., \& {Anderson}, S.~F. 2015, \apj, 802, 135

\bibitem[{{Ruszkowski} {et~al.}(2017){Ruszkowski}, {Yang}, \&
  {Reynolds}}]{yang17}
{Ruszkowski}, M., {Yang}, H.-Y.~K., \& {Reynolds}, C.~S. 2017, \apj, 844, 13

\bibitem[{{Sarazin} {et~al.}(2000){Sarazin}, {Irwin}, \&
  {Bregman}}]{sarazinetal00}
{Sarazin}, C.~L., {Irwin}, J.~A., \& {Bregman}, J.~N. 2000, \apjl, 544, L101

\bibitem[{{Scannapieco} \& {Oh}(2004)}]{s&o04}
{Scannapieco}, E., \& {Oh}, S.~P. 2004, \apj, 608, 62

\bibitem[{{Scannapieco} {et~al.}(2008){Scannapieco}, {Thacker}, \&
  {Couchman}}]{scannapiecoetal08}
{Scannapieco}, E., {Thacker}, R.~J., \& {Couchman}, H.~M.~P. 2008, \apj, 678,
  674

\bibitem[{{Schawinski} {et~al.}(2007){Schawinski}, {Thomas}, {Sarzi},
  {Maraston}, {Kaviraj}, {Joo}, {Yi}, \& {Silk}}]{schawinskietal07}
{Schawinski}, K., {Thomas}, D., {Sarzi}, M., {et~al.} 2007, \mnras, 382, 1415

\bibitem[{{Shakura} \& {Sunyaev}(1973)}]{s&s73}
{Shakura}, N.~I., \& {Sunyaev}, R.~A. 1973, \aap, 24, 337

\bibitem[{{Shankar} {et~al.}(2014){Shankar}, {Guo}, {Bouillot}, {Rettura},
  {Meert}, {Buchan}, {Kravtsov}, {Bernardi}, {Sheth}, {Vikram}, {Marchesini},
  {Behroozi}, {Zheng}, {Maraston}, {Ascaso}, {Lemaux}, {Capozzi},
  {Huertas-Company}, {Gal}, {Lubin}, {Conselice}, {Carollo}, \&
  {Cattaneo}}]{shankar14}
{Shankar}, F., {Guo}, H., {Bouillot}, V., {et~al.} 2014, \apjl, 797, L27

\bibitem[{{Silk} \& {Rees}(1998{\natexlab{a}})}]{SilkRees1998A&A}
{Silk}, J., \& {Rees}, M.~J. 1998{\natexlab{a}}, \aap, 331, L1

\bibitem[{{Silk} \& {Rees}(1998{\natexlab{b}})}]{s&r98}
---. 1998{\natexlab{b}}, \aap, 331, L1

\bibitem[{{Sivakoff} {et~al.}(2003){Sivakoff}, {Sarazin}, \&
  {Irwin}}]{sivakoffetal03}
{Sivakoff}, G.~R., {Sarazin}, C.~L., \& {Irwin}, J.~A. 2003, \apj, 599, 218

\bibitem[{{Spacek} {et~al.}(2016){Spacek}, {Scannapieco}, {Cohen}, {Joshi}, \&
  {Mauskopf}}]{spaceketal16}
{Spacek}, A., {Scannapieco}, E., {Cohen}, S., {Joshi}, B., \& {Mauskopf}, P.
  2016, \apj, 819, 128

\bibitem[{{Spergel} {et~al.}(2007){Spergel}, {Bean}, {Dor{\'e}}, {Nolta},
  {Bennett}, {Dunkley}, {Hinshaw}, {Jarosik}, {Komatsu}, {Page}, {Peiris},
  {Verde}, {Halpern}, {Hill}, {Kogut}, {Limon}, {Meyer}, {Odegard}, {Tucker},
  {Weiland}, {Wollack}, \& {Wright}}]{spergel07}
{Spergel}, D.~N., {Bean}, R., {Dor{\'e}}, O., {et~al.} 2007, \apjs, 170, 377

\bibitem[{{Springel} {et~al.}(2005){Springel}, {Di Matteo}, \&
  {Hernquist}}]{Springeletal2005MNRAS}
{Springel}, V., {Di Matteo}, T., \& {Hernquist}, L. 2005, \mnras, 361, 776

\bibitem[{{Steffen} {et~al.}(2006){Steffen}, {Strateva}, {Brandt}, {Alexander},
  {Koekemoer}, {Lehmer}, {Schneider}, \& {Vignali}}]{steffen06}
{Steffen}, A.~T., {Strateva}, I., {Brandt}, W.~N., {et~al.} 2006, \aj, 131,
  2826

\bibitem[{{Strickland} {et~al.}(2002){Strickland}, {Heckman}, {Weaver},
  {Hoopes}, \& {Dahlem}}]{stricklandetal02}
{Strickland}, D.~K., {Heckman}, T.~M., {Weaver}, K.~A., {Hoopes}, C.~G., \&
  {Dahlem}, M. 2002, \apj, 568, 689

\bibitem[{{Thacker} {et~al.}(2009){Thacker}, {Scannapieco}, {Couchman}, \&
  {Richardson}}]{thackeretal09}
{Thacker}, R.~J., {Scannapieco}, E., {Couchman}, H.~M.~P., \& {Richardson}, M.
  2009, \apj, 693, 552

\bibitem[{{Tombesi} {et~al.}(2015){Tombesi}, {Mel{\'e}ndez}, {Veilleux},
  {Reeves}, {Gonz{\'a}lez-Alfonso}, \& {Reynolds}}]{cr15}
{Tombesi}, F., {Mel{\'e}ndez}, M., {Veilleux}, S., {et~al.} 2015, \nat, 519,
  436

\bibitem[{{Tremaine} {et~al.}(2002){Tremaine}, {Gebhardt}, {Bender}, {Bower},
  {Dressler}, {Faber}, {Filippenko}, {Green}, {Grillmair}, {Ho}, {Kormendy},
  {Lauer}, {Magorrian}, {Pinkney}, \& {Richstone}}]{tremaineetal02}
{Tremaine}, S., {Gebhardt}, K., {Bender}, R., {et~al.} 2002, \apj, 574, 740

\bibitem[{{Trinchieri} \& {Fabbiano}(1985)}]{t&f85}
{Trinchieri}, G., \& {Fabbiano}, G. 1985, \apj, 296, 447

\bibitem[{{van de Voort} {et~al.}(2011){van de Voort}, {Schaye}, {Booth},
  {Haas}, \& {Dalla Vecchia}}]{v&f11}
{van de Voort}, F., {Schaye}, J., {Booth}, C.~M., {Haas}, M.~R., \& {Dalla
  Vecchia}, C. 2011, \mnras, 414, 2458

\bibitem[{{Verdier} {et~al.}(2016){Verdier}, {Melin}, {Bartlett}, {Magneville},
  {Palanque-Delabrouille}, \& {Y{\`e}che}}]{verdieretal16}
{Verdier}, L., {Melin}, J.-B., {Bartlett}, J.~G., {et~al.} 2016, \aap, 588, A61

\bibitem[{{Vernaleo} \& {Reynolds}(2006)}]{vernaleo06}
{Vernaleo}, J.~C., \& {Reynolds}, C.~S. 2006, \apj, 645, 83

\bibitem[{{Vernaleo} \& {Reynolds}(2007)}]{vernaleo07}
---. 2007, \apj, 671, 171

\bibitem[{{Vogt} {et~al.}(2015){Vogt}, {Dopita}, {Borthakur},
  {Verdes-Montenegro}, {Heckman}, {Yun}, \& {Chambers}}]{verdes15}
{Vogt}, F.~P.~A., {Dopita}, M.~A., {Borthakur}, S., {et~al.} 2015, \mnras, 450,
  2593

\bibitem[{{Vrtilek}(2000)}]{vrtilek00}
{Vrtilek}, J. 2000, {The Intragroup Medium: Distribution of Heavy Elements and
  Interaction with Radio Jets}, Chandra Proposal

\bibitem[{{Wagner} {et~al.}(2012){Wagner}, {Bicknell}, \& {Umemura}}]{wagner12}
{Wagner}, A.~Y., {Bicknell}, G.~V., \& {Umemura}, M. 2012, \apj, 757, 136

\bibitem[{{White} \& {Frenk}(1991)}]{w&f91}
{White}, S.~D.~M., \& {Frenk}, C.~S. 1991, \apj, 379, 52

\bibitem[{Yan {et~al.}(2011)Yan, Ho, Newman, Coil, Willmer, Laird, Georgakakis,
  Aird, Barmby, Bundy, Cooper, Davis, Faber, Fang, Griffith, Koekemoer, Koo,
  Nandra, Park, Sarajedini, Weiner, \& Willner}]{renbin11}
Yan, R., Ho, L.~C., Newman, J.~A., {et~al.} 2011, The Astrophysical Journal,
  728, 38

\bibitem[{{Yang} \& {Reynolds}(2016{\natexlab{a}})}]{yang2016}
{Yang}, H.-Y.~K., \& {Reynolds}, C.~S. 2016{\natexlab{a}}, \apj, 829, 90

\bibitem[{{Yang} \& {Reynolds}(2016{\natexlab{b}})}]{yang16}
---. 2016{\natexlab{b}}, \apj, 818, 181

\bibitem[{{Zanni} {et~al.}(2005){Zanni}, {Murante}, {Bodo}, {Massaglia},
  {Rossi}, \& {Ferrari}}]{zanni05}
{Zanni}, C., {Murante}, G., {Bodo}, G., {et~al.} 2005, \aap, 429, 399

\end{thebibliography}
\end{document}